\documentclass[prd,english,aps,showpacs,nofootinbib,preprint,eqsecnum]{revtex4-1}\usepackage[margin=2.5cm]{geometry}
\usepackage{amsmath,bm}
\usepackage{graphicx}
\usepackage{enumerate}
\usepackage{hhline}
\usepackage{graphicx,color}
\begin{document}

\title{Einstein Static Universe in Exponential $f(T)$ Gravity}

\author{Jung-Tsung Li}

\email{s100022519@m100.nthu.edu.tw}

\author{Chung-Chi Lee}

\email{g9522545@oz.nthu.edu.tw}

\author{Chao-Qiang Geng}

\email{geng@phys.nthu.edu.tw}

\affiliation{Department of Physics, National Tsing Hua University, Hsinchu 300,
Taiwan\\
Physics Division, National Center for Theoretical Sciences, Hsinchu
300, Taiwan}

\date{\today}

\begin{abstract}
We analyze the stability of the Einstein static closed and open universe in two types of exponential $f(T)$ gravity theories.
We show that the stable solutions exist in these two models. In particular, we find that  large regions of  parameter space 
in  equation of state $\mathsf{w}=p/\rho$ for the stable universe are allowed in the $f(T)$ theories.
\end{abstract}

%\pacs{04.50.Kd,04.20.Jb,04.25.Nx}
\maketitle

\section{Introduction}
\label{sec:introduction}
The Einstein static universe has recently been revived since our universe might evolve  from it
 to inflation. This is called the emergent universe with an  inflationary singularity to avoid a big bang singularity~\cite{Ellis 2004 1}.

One way to look at the theory beyond general relativity (GR) is  the teleparallel equivalence of general 
relativity (TEGR)~\cite{F.W. Hehl, K. Hayashi, E.E. Flanagan, J. Garecki,Geng:2011aj} introduced first  
by Albert Einstein~\cite{A. Einstein}.  Different from GR with the Levi-Civita connection,
 teleparallel gravity (TG) uses the Weitzenbock connection, which has no curvature (R) but has torsion (T). 
 To explain the late-time acceleration of the universe, a modified gravity
 has been proposed by extending T in the TG action  to an arbitrary function $f(T)$.
 % as an alternative to the inflation model. 
 In $f(T)$ gravity, torsion is responsible for the acceleration of the universe~\cite{Bengochea:2008gz, Linder:2010py}.
This modified gravity  would not only avoid the big bang singularity~\cite{Ferraro:2011zb} but also provide an alternative to inflation~\cite{Ferraro:2006jd}.
However, $f(T)$ gravity has some intrinsic problems, such as the violation of local Lorentz invariance~\cite{B. Li}.
In some of modified  gravity models, the solutions of the stable Einstein static universe do exist under linear homogeneous scalar perturbations as shown in
Refs.~\cite{Barrow:2003ni, Zhang:2010qwa}.
%, 
In Ref.~\cite{Wu:2011xa}, the stability of the Einstein static universe in a power law $f(T)$ plus the cosmological constant $\Lambda$ is studied. 
However,  $\Lambda$ is at the pre-inflation scale, which may lead to the hierarchy problem in the current universe.
To avoid this problem, we examine the stability of the Einstein static universe in two kinds of exponential $f(T)$ gravity 
theories~\cite{Linder:2010py}, which could explain the static Einstein universe in the pre-inflation era and be reduced back to TEGR in the present universe.

The paper is organized as follows. In Sec.~\ref{sec:fTgravity}, we  briefly introduce  the formulation of  $f(T)$ gravity
with the Weitzenbock connection. We also show two explicit forms of the exponential theories.
In Sec.~\ref{sec: einstein static universe},  we give the conditions for the stable Einstein static solutions.
In Sec.~\ref{sec: exponential gravity}, we study the stable solutions in the exponential models.
We illustrate the oscillating behavior of the Einstein universe in  Sec.~\ref{sec: numerical}.
%, the numerical study of the dynamics of the  in Type I Exponential Model has been done.
%
%Finally, 
We present our conclusions  in Sec.~\ref{sec:Summary}.

\section{Formulation in $f(T)$ Gravity}
\label{sec:fTgravity}
In teleparallelism, the dynamical object is the vierbein field $\bm{e}_i(x^\mu)$, which is an orthogonal basis for the tangent space at the point $x^\mu$ of the manifold
with the relation: $\bm{e}_i \cdot \bm{e}_j= \eta_{ij}$, where $\eta_{ij}=$diag(1,-1,-1,-1). The metric tensor is given by
\begin{equation}
g_{\mu\nu}=\eta_{ij}e_\mu^i(x) e_\nu^j(x)\,
 \end{equation}
with
\begin{equation}
e_\mu^i(x) e^\mu_j(x)=\delta^i_j\,.
\end{equation}
In this formulation, the Weitzenbock connection is used and
the torsion tensor is defined by
\begin{equation}
T^\lambda_{\mu\nu}=\Gamma^\lambda_{\nu\mu}-\Gamma^\lambda_{\mu\nu}  =e^\lambda_i(\partial_\mu e^i_\nu-\partial_\nu e^i_\mu)\,,
\end{equation}
where
\begin{equation}
\Gamma^\lambda_{\mu\nu}=e^\lambda_i \partial_\nu e^i_\mu\,.
\end{equation}
The action in TG is expressed as
\begin{equation}
\label{eq2.8}
I=\frac{1}{16\pi G} \int eT d^4 x\,,
\end{equation}
where $e\equiv det (e^i_\mu)=\sqrt{-g}$ and $T$ is the torsion scalar, defined by
\begin{equation}
T=S_\lambda^{\enspace\mu\nu}T^\lambda_{\enspace\mu\nu}\,,
\end{equation}
with
\begin{equation}
S_\lambda^{ \enspace\mu\nu} \equiv  \frac{1}{2}(K^{\mu\nu}_{\quad\lambda} + \delta^\mu_{\enspace\lambda}T^{\theta\nu}_{\quad\theta}-\delta^\nu_{\enspace\lambda}T^{\theta\mu}_{\quad\theta})\,,
\end{equation}
and 
\begin{equation}
K^{\mu\nu}_{\quad\lambda} =\frac{-1}{2}(T^{\mu\nu}_{\quad\lambda}-T^{\nu\mu}_{\quad\lambda}-T_\lambda^{\enspace\mu\nu})\,.
\end{equation}
The modified teleparallel action for $f(T)$ gravity 
is given by~\cite{Linder:2010py} 
\begin{equation}
I=\frac{1}{16\pi G} \int ef(T) d^4 x\,,
\label{FT}
\end{equation}
where $f(T)$ is an arbitrary function of $T$.
In this paper, we will concentrate on the following two types of  the exponential $f(T)$ models:

\begin{equation} 
f(T)=T+\alpha T\left(1-e^{\beta T_0/T}\right),
\label{eq: exp I}
\end{equation}
and
\begin{equation} f(T)=T+\alpha T_0\left(1-e^{\beta T^2/T_0^2}\right),
\label{eq: exp II}
 \end{equation}
where $\alpha$, $\beta$ and $T_0$ are constants.
The motivation of these two  models is that  $f(T)$ can be reduced back to TEGR for small $T$, corresponding to the current universe. 
Explicitly,
when $T$ is very small compared to $T_0$ and $\beta <0$, 
the two models in Eqs.~(\ref{eq: exp I}) and (\ref{eq: exp II}) give
%can get back to TEGR, i.e. 
$f(T)\approx(1+\alpha)T$ and
%, while the second one in Eq.~(\ref{eq: exp II}) becomes to 
$f(T)\approx T$, respectively.

\section{Einstein Static Universe}
\label{sec: einstein static universe}
%\label{sec:StaticUniverse}
%
We now consider the FRW metric
\begin{equation}
ds^2=dt^2-k^2a^2(t)[d(k\psi)^2+\sin^2(k\psi)(d\theta ^2+\sin^2\theta d\phi^2)]
\end{equation}
where its vierbein fields are given by~\cite{Ferraro:2011us}
\begin{eqnarray}
e^0_0=&&1, \quad e^0_\psi=e^0_\theta=e^0_\phi=e^1_0=e^2_0=e^3_0=0,\nonumber\\
e^1_\psi=&&-a(t) k^2\cos\theta,\quad e^1_\theta=a(t)k \sin(k\psi) \sin\theta\cos(k\psi),\quad e^1_\phi=-a(t)k\sin^2(k\psi)\sin^2\theta,  \nonumber\\
e^2_\psi=&&a(t)k^2 \sin\theta \cos\phi,\quad e^2_\theta=-a(t)k \sin(k\psi) \left[\sin(k\psi) \sin\phi-\cos(k\psi) \cos\theta \cos\phi \right], \nonumber\\
e^2_\phi=&&-a(t)k \sin(k\psi)\sin\theta \left[ \cos(k\psi) \sin\phi+\sin(k\psi) \cos\theta \cos\phi \right],
 \nonumber\\
e^3_\psi=&&-a(t)k^2\sin\theta \sin\phi,\quad e^3_\theta=-a(t)k\sin(k\psi) \left[\sin(k\psi) \cos\phi +\cos(k\psi) \cos\theta \sin\phi \right], \nonumber \\
e^3_\phi=&&-a(t)k\sin(k\psi) \sin\theta \left[ \cos(k\psi) \cos\phi-\sin(k\psi) \cos\theta \sin\phi \right]  ,
\end{eqnarray}
with $k = 1$ and $i$ or $k^2=\pm1$, representing the closed and open universe, respectively. The torsion scalar can be written as
\begin{equation}
T=6( k^{2} a^{-2} -H^2)\,.
\end{equation}
As a result, the modified Friedmann equation is given by~\cite{Ferraro:2011us}
\begin{equation}
12H^2f'(T)+f(T)=16\pi G\rho \equiv \kappa \rho,
\label{eq: friedmann 1}
\end{equation}
\begin{equation}
(k^2 a^{-2}+\dot H)(48H^2f''(T)+4f'(T))-f'(T)(8\dot H+12H^2)-f(T)=\kappa p\,.
\label{eq: friedmann 2}
\end{equation}
To get an Einstein static universe,  the conditions of $\dot a=H=0$, $\ddot a=0$, and $T_0=T(a_0)=k^2 6/a_0^2$
are imposed.
By using Eqs. (\ref{eq: friedmann 1}) and (\ref{eq: friedmann 2}), we obtain 
\begin{equation}
f_0\equiv f(T_0)=\kappa \rho _0\,,
\label{eq: constraint 1}
\end{equation}
\begin{equation} k^2  \frac{4f'_0}{a_0^2}-f_0=\kappa p_0\,,
\label{eq: constraint 2}
\end{equation}
with $f'_0\equiv df/dT\lvert _{T=T_0}$, $\rho_0=\rho (a_0)$ and $p_0=p(a_0)$.
By combining Eqs.~(\ref{eq: constraint 1}) and (\ref{eq: constraint 2}), we find
\begin{equation}
\left(\frac{Tf'}{f}\right)_{T=T_0}=\frac{3}{2}\left(1+\mathsf{w}\right)\,,
\label{eq: costraint 3}
\end{equation}
where $\mathsf{w}=p/\rho$ is  equation of state of the background matter.

We now perform the linear homogeneous scalar perturbations in the static Einstein universe. The perturbations in $a$ and $\rho $ depend only on time,
$i.e.$,
\begin{equation}
a(t)=a_0(1+\delta a(t)), \quad \rho(t)=\rho _0(1+\delta \rho (t))\,,
\label{eq: perturbation}
\end{equation}
with~\cite{Wu:2011xa}
\begin{equation}
\delta a(t)=C_1e^{\gamma t}+C_2e^{-\gamma t}\,.
\label{eq: oscillation}
 \end{equation}
By using the procedure in Ref.~\cite{Wu:2011xa}, we get
\begin{equation}
\gamma^2=\frac{\kappa \rho _0}{4f'_0{^2}}(1+\mathsf{w} )\left[\left(1+3\mathsf{w} \right)f'_0-3\left(1+\mathsf{w} \right) \frac{f_0f''_0}{f'_0}\right].
\label{eq: gamma square}
\end{equation}
 From Eq.~(\ref{eq: oscillation}), we obtain an oscillating  universe if $\gamma^2 < 0$, corresponding to the stable Einstein static universe.

In the following section, we discuss the stable Einstein static solutions of the exponential $f(T)$ models in Eqs.~(\ref{eq: exp I}) and~(\ref{eq: exp II}).
% with closed and open universe
%

\section{Conditions for Stable Einstein Static Solutions }
\label{sec: exponential gravity}
\subsection{The first type of the exponential models}
In the closed universe, we have $T_0=6a_0^{-2}$ for the Einstein static solution. 
Substituting Eq.~(\ref{eq: exp I}) into Eqs.~(\ref{eq: constraint 1}) and (\ref{eq: constraint 2}), 
we obtain the constraints for the Einstein static universe
\begin{equation}
T_0(1+\alpha -\alpha e^\beta)=\kappa \rho _0\,,
\label{eq: I-closed-constraint-1}
\end{equation}
\begin{equation}
\frac{2}{3}T_0\alpha\beta e^\beta=\kappa \rho_0\left(\mathsf{w} +\frac{1}{3}\right)\,,
\label{eq: I-closed-constraint-2}
\end{equation}
leading to 
\begin{equation}
\frac{T_0}{\kappa\rho_0}=1-\frac{1-e^\beta}{2\beta e^\beta}\left(1+3\mathsf{w}\right)>0\,,
\label{eq: constraint exp closed}
\end{equation}
where $T_0/(\kappa\rho_0)$ is positive since $\rho_0$ is positive definite and
$T_0>0$ in the closed universe. 

\begin{figure}[h]
\begin{minipage}[t]{0.45\linewidth}
\includegraphics[width=\linewidth]{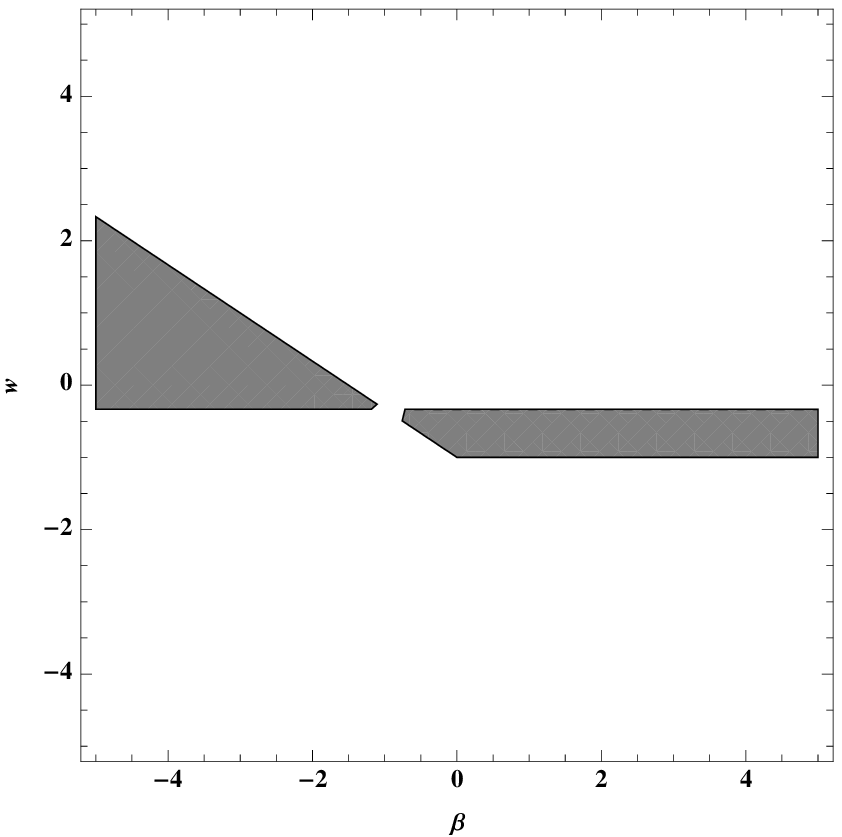}
\end{minipage}
\begin{minipage}[t]{0.45\linewidth}
\includegraphics[width=\linewidth]{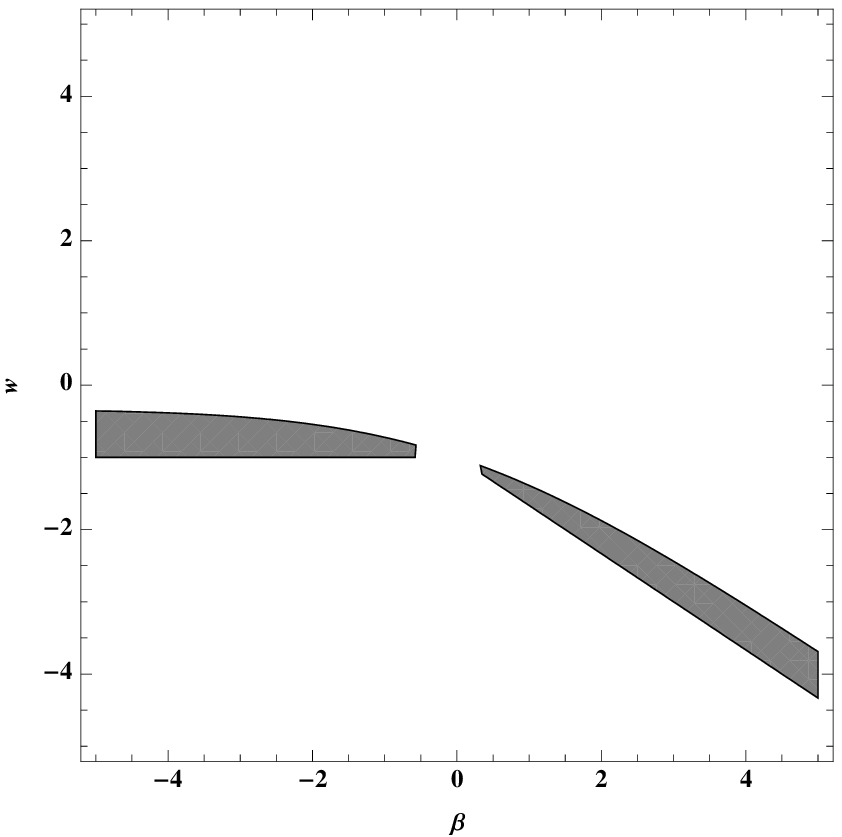}
\end{minipage}
\caption{Stable solutions (shaded regions) of the first exponential gravity model in the spatially closed (left panel) and open (right panel) Einstein static universe.}
\label{fig: exp I}
\end{figure}

 From Eqs.~(\ref{eq: exp I}), (\ref{eq: gamma square}), (\ref{eq: I-closed-constraint-1}) and (\ref{eq: I-closed-constraint-2}), we have
\begin{equation}
\label{eq: close exp gamma}
\gamma^2=\frac{3\kappa^2\rho_0^{2}}{8T_{0}f^{'2}_0}\left(1+\mathsf{w}\right)\left(1+3\mathsf{w}\right)\left[\left(1+\frac{2\beta}{3}\right)+\mathsf{w}\right].
\end{equation}
Note that $f(T) \to T+\alpha\, \beta\,T_0$ when   $\beta\to 0$. If the combination $\alpha\, \beta\,T_0$ is finite, we may define it as the cosmological constant $\Lambda$.
As expected,  in the limit of $f(T) \to T+\Lambda$, one has 
\begin{equation}
\gamma^2=\frac{\kappa\rho_0}{4}(1+\mathsf{w})(1+3\mathsf{w}), \qquad  \Lambda=\frac{\kappa\rho_0}{2}(1+3\mathsf{w})\,,
\end{equation}
which is the result in GR.
For the stable universe in GR, which requires $\gamma^2<0$, one finds
\begin{equation}
\label{eqEoS}
-1<\mathsf{w}<-{1\over 3}\,.
\end{equation}
In Fig.~\ref{fig: exp I} (left panel), 
 we show the stable regions ($\gamma^2<0$) based on Eqs.~(\ref{eq: constraint exp closed}) and (\ref{eq: close exp gamma}).
For $\beta=0$,  as seen from the figure, we obtain the same allowed region of $-1<\mathsf{w}<-1/3$ as that in Eq.~(\ref{eqEoS}).

In the case of the open universe, 
%in which $T_0=-6a_0^{-2}$ for the Einstein static solution,
%
from Eqs.~(\ref{eq: constraint 1}) and (\ref{eq: constraint 2}) we get the constraints for the Einstein open static universe as:
\begin{equation}
T_0(1+\alpha -\alpha e^\beta)=\kappa \rho_0\,,
\label{eq: constraint open 1}
\end{equation}
\begin{equation}
\frac{2}{3}T_0\alpha\beta e^\beta=\kappa \rho_0\left(\mathsf{w} +\frac{1}{3}\right)\,,
\label{eq: constraint open 2}
\end{equation}
respectively, where $T_0=-6a_0^{-2}$.
With these two constraints, we find
\begin{equation}
\frac{T_0}{\kappa\rho_0}=1-\frac{1-e^\beta}{2\beta e^\beta}\left(1+3\mathsf{w}\right)\,,
\label{eq: constraint open 3}
\end{equation}
which is negative because $T_0=-6a_0^{-2}<0$ in the static open universe. From $f(T)$ in Eq.~(\ref{eq: gamma square}), we derive
\begin{equation}
\gamma^2=\frac{3\kappa^2 \rho_0^2}{8T_0f_0^{'2}}(1+\mathsf{w})(1+3\mathsf{w})\left(1+\frac{2\beta}{3}+\mathsf{w}\right)\,.
\label{eq: open exp gamma}
\end{equation}
In Fig.~\ref{fig: exp I} (right panel), we display the  stable solutions ($\gamma^2<0$) for the Einstein static open universe.
Clearly,  there is no stable region at $\beta=0$. Thus, the universe is not stable in the spatially open universe.

\subsection{The second Type of the exponential models}
\begin{figure}
\begin{minipage}[t]{0.45\linewidth}
\includegraphics[width=\linewidth]{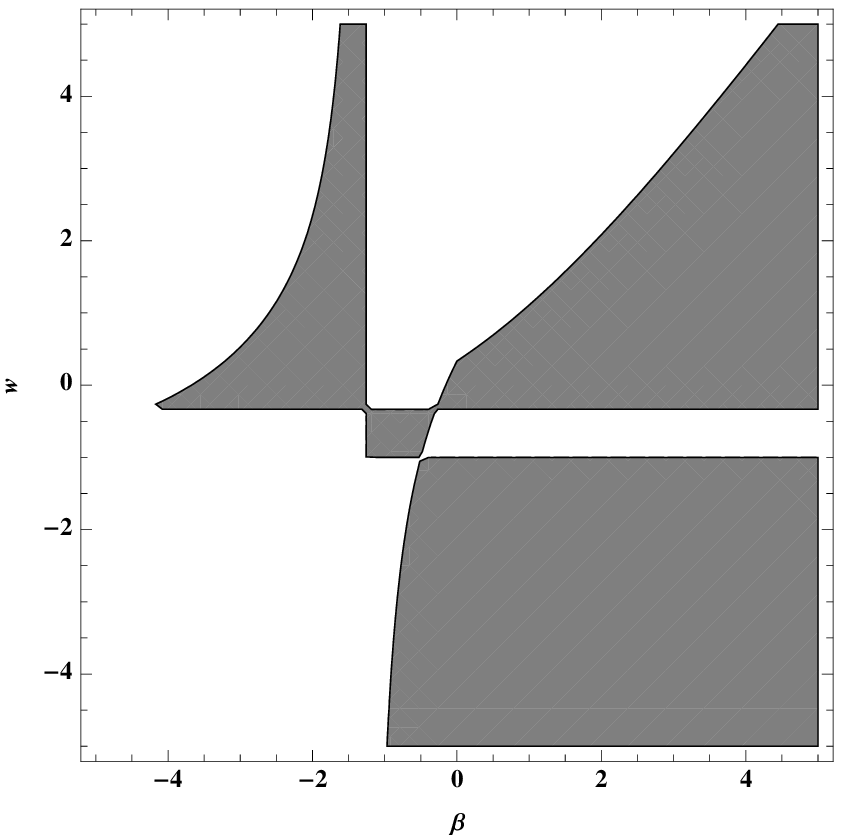}
\end{minipage}
\begin{minipage}[t]{0.45\linewidth}
\includegraphics[width=\linewidth]{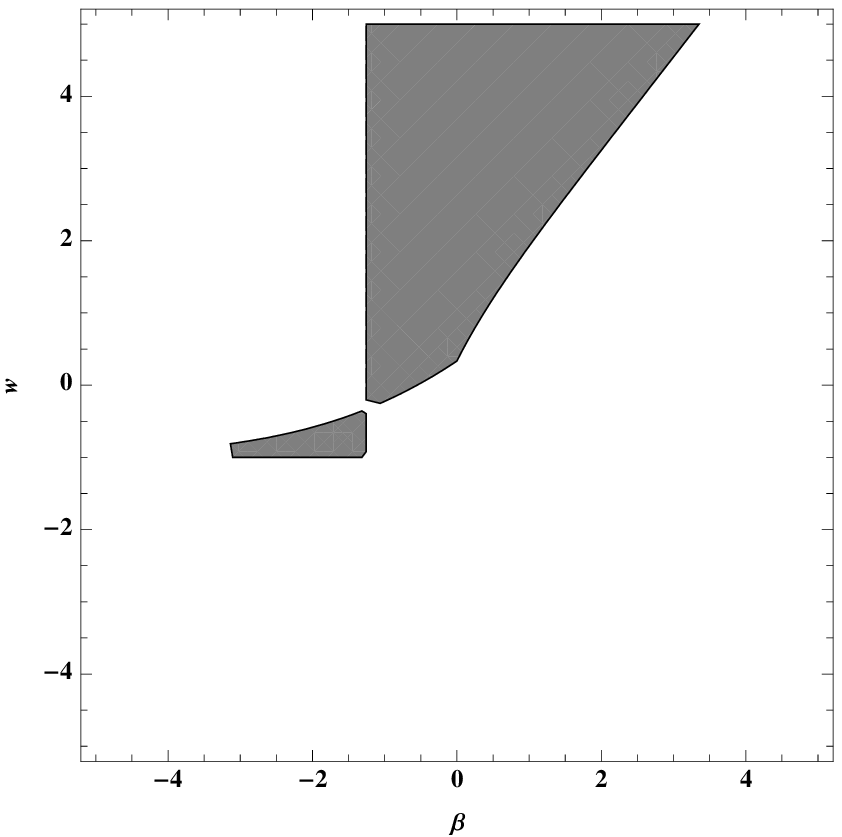}
\end{minipage}
\caption{Legend is the same as Fig.~\ref{fig: exp I} but for the second model.}
\label{fig: exp II}
\end{figure}

For the second type of the exponential models in Eq.~(\ref{eq: exp II}), we obtain conditions for the closed Einstein static universe as 
\begin{equation} f_0=T_0(1+\alpha-\alpha e^{\beta})=\kappa \rho_0,
\end{equation}
\begin{equation} \frac{2}{3}T_0(1-2\alpha \beta e^{\beta})-\kappa\rho_0=\kappa p_0,
\end{equation}
leading to 
\begin{equation}\frac{T_0}{\kappa\rho_0}=\frac{\frac{3}{2}(1+\mathsf{w})(1-e^{\beta})+2\beta e^{\beta}}{1-e^{\beta}+2\beta e^{\beta}},
\label{eq: constraint exp II}
\end{equation}
which will be positive because $T_0=6a_0^{-2}>0$ in the closed universe.
Substituting Eq.~(\ref{eq: exp II}) into Eq.~(\ref{eq: gamma square}), we get
\begin{equation}
\gamma^2=\frac{3\kappa^2\rho_0^2}{8T_0f_0^{'2}}(1+\mathsf{w})\left[(1+\mathsf{w})(1+3\mathsf{w})
-2(1+2\beta)\left(\mathsf{w}+1-\frac{(1+\mathsf{w})(1-e^{\beta})+\frac{4}{3}\beta e^{\beta}}{1-e^{\beta}+2\beta e^{\beta}}\right)\right]\,.
\label{eq: closed exp II gamma}
\end{equation}
  From Eqs.~(\ref{eq: constraint exp II}) and (\ref{eq: closed exp II gamma}), we can plot the stable solutions ($\gamma^2<0$) of 
the Einstein static closed universe in Fig.~\ref{fig: exp II} (left panel).

In the open universe, the forms of equations are the same as those in the closed universe. 
However,  since $T_0=-6a_0^{-2}<0$ in the static open universe,  the constraint in Eq.~(\ref{eq: constraint exp II-2}) is negative, $i.e.$
\begin{equation}
\frac{T_0}{\kappa\rho_0}=\frac{\frac{3}{2}(1+\mathsf{w})(1-e^{\beta})+2\beta e^{\beta}}{1-e^{\beta}+2\beta e^{\beta}}<0,
\label{eq: constraint exp II-2}
\end{equation}
while $\gamma^2$ is also negative for the stable static universe: 
\begin{equation}
\gamma^2=\frac{3\kappa^2\rho_0^2}{8T_0f_0^{'2}}(1+\mathsf{w})\left[(1+\mathsf{w})(1+3\mathsf{w})-2(1+2\beta)\left(\mathsf{w}+1-\frac{(1+\mathsf{w})(1-e^{\beta})+\frac{4}{3}\beta e^{\beta}}{1-e^{\beta}+2\beta e^{\beta}}\right)\right].
\label{eq: closed exp II gamma-2}
\end{equation}
The stable solutions for the open universe are illustrated 
in Fig.~\ref{fig: exp II} (right panel).
% Note that $T_0=-6a_0^{-2}$ in Eq.~(\ref{eq: closed exp II gamma-2}) is negative.
%

\section{Oscillating Einstein Universe}
\label{sec: numerical}
 From  Eq.~(\ref{eq: oscillation}), it is obvious that the stable Einstein static universe is oscillating. 
 The behavior has also been explicitly demonstrated in other gravity theories,  such as
the vacuum energy model~\cite{Carneiro:2009et} and DGP braneworld scenario~\cite{Zhang:2010qwa}. 
In this section, we would like to give similar discussions to illustrate this oscillating
 behavior in  the first type of the exponential models in Eq.~(\ref{eq: exp I}). The results can
be easily extended to the second one.

\begin{figure}[tbp]
\begin{minipage}[t]{0.45\linewidth}
\includegraphics[width=\linewidth]{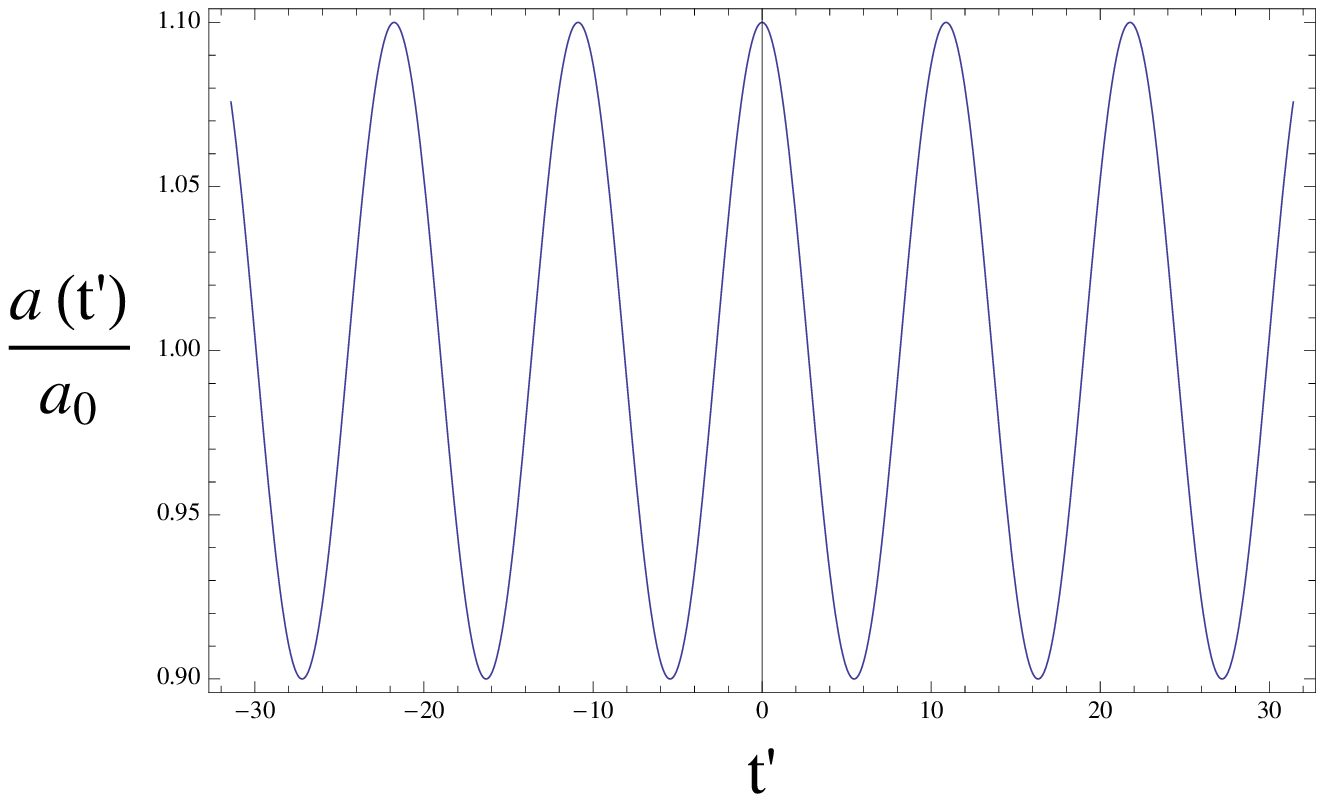}
\end{minipage}
\begin{minipage}[t]{0.45\linewidth}
\includegraphics[width=\linewidth]{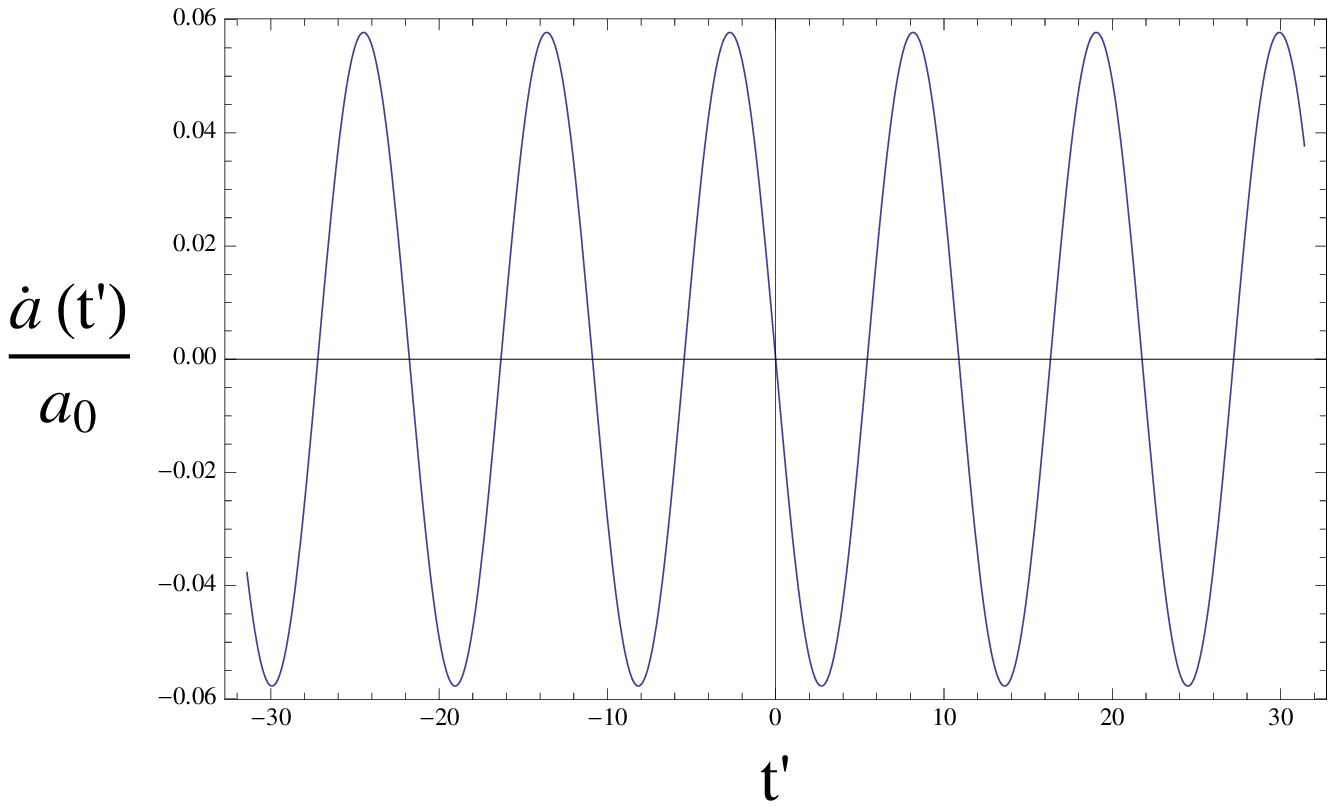}
\end{minipage}
\caption{$a(t^{\prime})$ (left panel) and  $\dot a(t^{\prime})$ (right panel)  as functions of $t^{\prime}( =\sqrt{ \frac{3\kappa^2\rho_0^2}{8T_0f_0'^2}} t)$
 in the first exponential gravity model in the
closed universe.}
\label{fig: type I closed osci}
\end{figure}

Since an obvious stable solution for the first exponential closed universe comes from
$\beta=-2$ and $\mathsf{w}=0$,
 we can get  
\begin{equation}
\gamma=\sqrt{\frac{3\kappa^2\rho_0^2}{8T_0f_0^{'2}}}\cdotp{i\over\sqrt3}
\end{equation}
from Eq.~(\ref{eq: close exp gamma}).
Consequently, the scale factor is given by
\begin{equation}
a(t)=a_0\left[1+C\sin\left(\sqrt{\frac{3\kappa^2\rho_0^2}{8T_0f_0^{'2}}}{t\over\sqrt3}+\eta\right) \right].
\end{equation}
By defining $t^{\prime} \equiv \sqrt{ \frac{3\kappa^2\rho_0^2}{8T_0f_0'^2}} t$,
we get
\begin{eqnarray}
a(t^{\prime})&=& a_0\left[1+C\sin\left({{t^{\prime}}\over\sqrt3}+\eta\right) \right],
\nonumber\\
\dot a(t^{\prime}) &=& {a_0 C \over\sqrt3}\cos\left({{t^{\prime}}\over\sqrt3}+\eta \right).
\end{eqnarray}
Using initial conditions given by $a(t=0)=1.1 \times a_0$ and $\dot a(t=0)=0$, we obtain $\eta=\pi/2$ and $C=0.1$.
We plot  the stable small oscillating closed universe for the first exponential gravity model in Fig.~\ref{fig: type I closed osci}.

\begin{figure}[tbp]
\begin{minipage}[t]{0.45\linewidth}
\includegraphics[width=\linewidth]{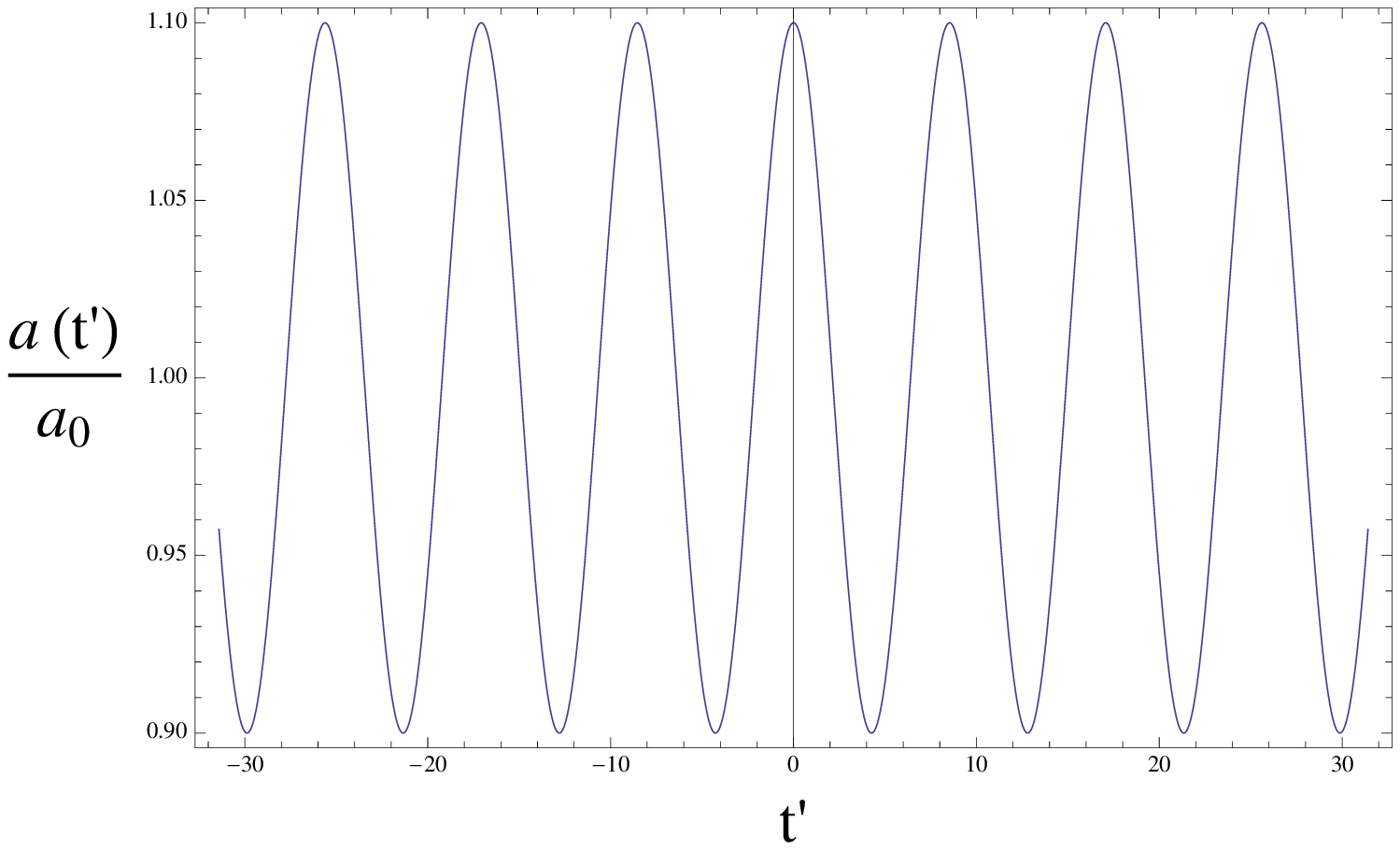}
\end{minipage}
\begin{minipage}[t]{0.45\linewidth}
\includegraphics[width=\linewidth]{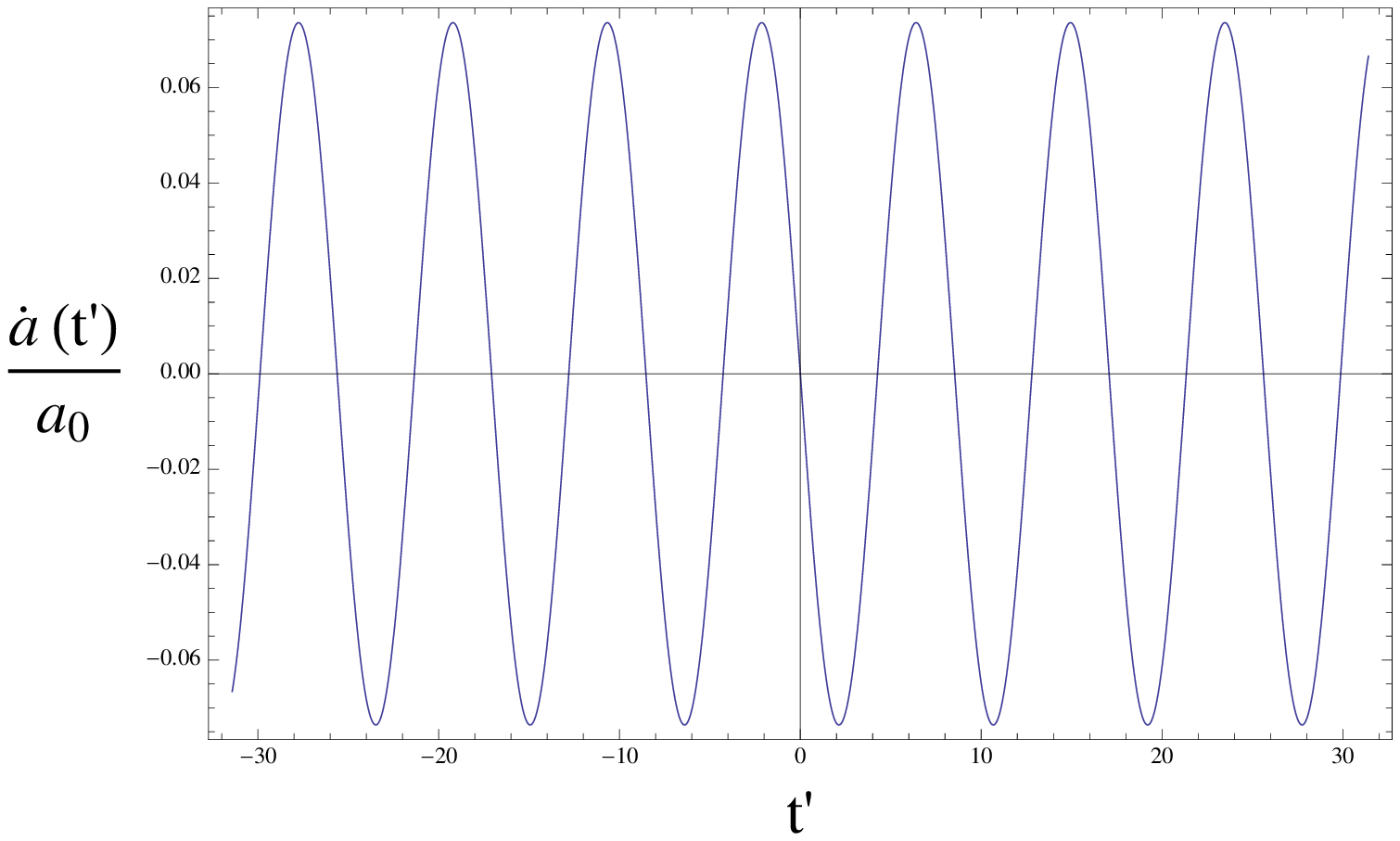}
\end{minipage}
\caption{Legend is the same as Fig.~\ref{fig: type I closed osci} but in the  open universe with 
$t^{\prime} =\sqrt{ \frac{3\kappa^2\rho_0^2}{8\lvert T_0\rvert f_0'^2}} t$.}
\label{fig: type I open osci}
\end{figure}

For the open universe, we pick the stable solution of
$\beta=-4$ and $\mathsf{w}=- 1/2$.
 With the same initial conditions as  the closed universe,
 we find
\begin{eqnarray}
\frac{a(t^{\prime})}{a_0}&=&\left[1+\frac{1}{10}\sin\left(\sqrt{\frac{13}{24}}t^{\prime}+{\pi\over2}\right)\right],
\nonumber\\
\frac{\dot a(t^{\prime})}{a_0} &=& \frac{1}{10} \sqrt{\frac{13}{24}}\cos\left(\sqrt{\frac{13}{24}}t^{\prime}+{\pi\over2}\right),
\end{eqnarray}
where $t^{\prime} =\sqrt{ \frac{3\kappa^2\rho_0^2}{8\lvert T_0\rvert f_0'^2}} t$.
In Fig.~\ref{fig: type I open osci}, we show the oscillation in the open case, which is not allowed in GR.

\section{Conclusions}
\label{sec:Summary}
We have discussed the linear homogeneous scalar perturbations near the Einstein static universe in $f(T)$ gravity. 
We have explicitly studied both closed and open universe in the two exponential $f(T)$ gravity models.
These two $f(T)$ models are proposed in order to explain the pre-inflation universe and be reduced back to TEGR in the present universe.
We have demonstrated that in these models, the Einstein static universe can be stable in both open and closed cases with  large allowed regions of
 $\mathsf{w}$. Note that in GR, only the closed Einstein universe contains stable solutions with  $-1<\mathsf{w}<-{1\over 3}$
for the background matter.
Explicitly, for the first exponential  gravity model of $f(T)=T+\alpha T(1-e^{\beta T_0/T})$,
we have shown that $\mathsf{w}>-1$  ($\mathsf{w}<-{1\over 3}$) in the closed (open) universe  can have the stable solution.
In the limit  $\beta \rightarrow 0$, 
%we have demonstrated that 
the model gives the GR result as
$f(T) \rightarrow T+\Lambda$ with $\Lambda=\alpha\beta T_0$ being the cosmological constant.
For the second exponential gravity model of $f(T)=T+\alpha T_0(1-e^{\beta T^2/T_0^2})$, 
we have found that $\mathsf{w}>-1$ (arbitrary $\mathsf{w}$)
is the stable solution for the open (closed) Einstein universe.
We have also illustrated  the oscillating behaviors of the stable Einstein static universe in the first model. 
We have shown that both $a(t)$ and $\dot a(t)$ are oscillating without divergences as expected.
\\

\begin{acknowledgments}
We are grateful to Professor Ling-Fong Li for reading the manuscript.
The work was supported in part by National Center for Theoretical Science
and  National Science Council (NSC-98-2112-M-007-008-MY3 and
NSC-101-2112-M-007-006-MY3) of R.O.C.
\end{acknowledgments}

\end{document}